# AnAmeter: The First Steps to Evaluating Adaptation

F. Tarpin-Bernard[1] , I. Marfisi-Schottman[1], H. Habieb-Mammar[2]

[1] LIESP
Université de Lyon, LIESP,
INSA-Lyon, F-69621, Villeurbanne, France
Tel: +33 (0) 4 72 43 70 99
Fax: +33 (0) 4 72 43 79 92
franck.tarpin-bernard@insa-lyon.fr
iza.marfisi@insa-lyon.fr

[2] Human-Centered Software Engineering Group
Concordia University, 1455 Maisonneuve West
Montreal Quebec Canada H3G 1M8
mammar@encs.concordia.ca

**Abstract.** Computer systems now need to be adaptable in order to meet the needs of ever widening populations using a large variety of interactive services and various computer devices in different technical and social environments. This paper presents the online AnAmeter framework that helps characterize the different types of adaptations a system features by helping the evaluator fill in a simple form. The provided information is then processed to obtain a quantitative evaluation of three parameters called global, semi-global and local adaptation degrees. By characterizing and quantifying adaptation, AnAmeter provides the first steps towards the evaluation of the quality of a system's adaptation. AnAmeter is an open tool available as freeware on the web and has been applied to a selection of well known systems.

## 1 Introduction

More and more developers are now being asked to design interactive systems which are compatible with a large diversity of users accessing various functionalities and information using a range of different computing platforms. People using computer systems are of various ages and have all different kinds of interests and background knowledge. In addition to traditional desktops, the variety of computing platforms includes mobile telephones, personal digital assistants (PDAs), pocket PCs, wearable and immersive environments and many more. In this context, novel user interface metaphors and interaction styles are emerging.

Faced with this huge set of propositions, it is very difficult to characterize to what extent a specific application is adaptable. Likewise, it is difficult to explicit the new adaptation features that should be implemented in this system in order to increase its adaptation degree. For these reasons, it is necessary to *characterize* all the different kinds of adaptations that can possibly exist and define a proper way of *quantifying* the degrees of these adaptations. In order to accomplish a good evaluation framework, a subjective measure of the user's satisfaction with the adaptation could be added (see Fig. 1). These indicators could be used either for identifying strengths and weaknesses of a system, or for objectively comparing several systems.

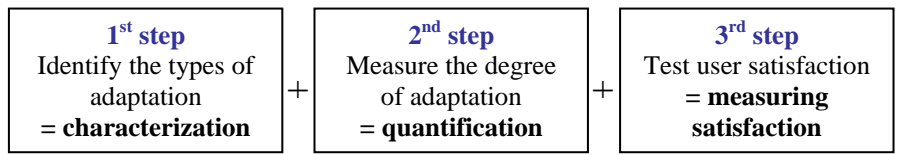

**Figure 1:** The three steps to evaluating a systems adaptation

In this paper, we present a first proposal, AnAmeter, for *characterizing* and *quantifying* the adaptation of a system. This tool is largely based on our analysis of the multiple facets of adaptation we will develop in the first section. Then, we present the core of AnAmeter: a grid that helps characterize the adaptations by crossing the adaptation factors and the aspects of adaptation. Based on this grid, we build a quantification technique that provides a measure of the adaptation degree. Finally, the interest, limits and potential extensions of AnAmeter are discussed.

## 2 The multiple Facets of adaptation

Many studies have tried to build a taxonomy of adaptive interfaces [1, 2, 3]. Based on these works and on a first design space provided by Vanderdonckt & al. [4], we consider that the most important questions to characterize adaptation are:

- ➢ Who initiates the adaptation?
- ➢ What are the factors to which the system can adapt?
- ➢ What aspects of the system are adaptable?

In the next part, we will look at each of these questions in detail and elaborate a list of possible answers.

### 2.1 Who initiates the adaptation?

After using Dieterich's [5] definition of adaptation, Kobsa & al. [6] identified two possible sources responsible of initiating the adaptation:
- The system its-self. In this case the adaptation is automatically initiated and the system is called *adaptive*.
- The user. In this case the adaptation is requested by the user and the system is called *adaptable*.

## 2.2 What are the factors to which the system can adapt?

Usually, the need for adaptation is associated to the notion of context. In the human computer interface (HCI) field, a context is generally described according to three dimensions along which adaptability can be analyzed [3]: the user, the platform and the environment. However, a specific user placed in the same environment using the same interaction platform could require some adaptation relevant to his/her activity. That is why, considering our goal, which is to find a way of characterizing and quantifying adaptability and adaptivity, we define four sets of factors of adaptation: user, interaction platform, environment and the activity.

In the next part, we will itemize these four factors of adaptation into sub-factors. The lists of sub-factors are not meant to be exhaustive, but as we will see later, they will help evaluate the levels of adaptation provided by a system.

This classification work was done thanks to a detailed reviews of the state of the art and "context of use" definitions provide by standards like IEC CDV TR 61997 [7] and ISO 9241-11 [8]. We also analyzed more than 50 systems found in articles or available in public distribution (complete list on http://liesp.insa-lyon.fr/AnAmeter/References.php).

### 2.2.1 User

A User model usually refers to various user characteristics [9, 10]. The user model contains characteristics of a particular user (also called stereotype). We can group these characteristics into four sub-factors:

Knowledge and level of experience: The knowledge of an individual user refers to the user's theoretical understanding of the subjects treated in the system. The level of experience refers to the skills acquired to use the system itself. Certain educational systems adapt the content of the lessons to the student's knowledge. Other systems give more helping tips to users that are not familiar with the system.

Socio-demographic characteristics and user role: Socio-demographic characteristics such as age, gender, weight, height, wage, profession, hobbies, cultural preferences… are becoming important factors of adaptation for all kinds of systems. When a user asks his GPS for tourist activities in a given area for example, the system will filter the information showing only the attractions compatible with the user's interests and his propensity to spend. In certain cases it is also worthwhile to consider the user's role. For example, a tutoring system can be presented in different ways to a student or to a teacher.

Cognitive abilities and emotional state: The cognitive abilities represent the ease with which the user deals with different modes of perceiving, memorizing, learning, judging etc. Introduced by Picard, R.W. "affective computing" [12], the emotional state (happy, sad, worried, frustrated, panicked, confident…) is also a very important characteristic to take into account when adapting a system.

Perceptual and motor abilities: These characteristics are useful to enable the systems to be used with disabilities (vision, manipulation, etc.). These disabilities can range from slight myopia or color blindness to total deathless and paralysis.

### 2.2.2 Interaction platform

The interaction platform describes the physical characteristics of the devices. The major characteristics that a system may take into account are the following:

Computing power and autonomy: Systems often need to be adapted to the platform's processing power and the memory capacity. For some portable devices it is also worthwhile to adapt to the battery level by shutting of certain services for example.

Input/Output device: Some systems are available on a wide variety of platforms. Certain web browsers, for instance, adapt to the different screen sizes and input devices such as mice, keyboards and pens when used on desktops, laptops, or telephones.

Software environment: Computer systems are almost always used alongside other systems on the same platform. These systems can adapt to cohabitate, synchronize and even cooperate with each other. The msn window, for example, can be configured to stay on top of other windows allowing msn to cohabitate with the other applications running at the same time.

Connectivity: More and more systems are now using network connections. The connectivity factor is therefore very important. Systems can adapt to cope with the lake of connection or slow connectivity but we may also want them to adapt to the type of network.

### 2.2.3 Environment

The third factor is the environment, a term used to cover the physical, social and organizational elements that are outside of the interactive system (platform & user).

Human environment: In some cases, systems can adapt to the other people who are interacting with the user (directly or thru the system). This kind of adaptation can be used for multi-user applications or for applications that detect humans present in the same physical area as the user and who are susceptible of communicating. A camera that automatically widens the view when a second person enters the room is an example of adaptation to the human environment.

Machine environment: This type of environment is defined by any reachable material such as webcams, printers, screens that could be connected on the fly to the main system, but which are not mandatory for the interactive system.

Ambient characteristics: It is very common to find systems that adapt their interface colors when night falls (e.g. GPS map systems) but the luminance is not the only aspect of ambient conditions that deserves attention. Systems can also adapt to the temperature, the level of noise and the movements of the device.

Spacio-temporal characteristics: Many GPS navigation systems propose potential interesting tourist areas by using geographic latitude and longitude measures. Localisation can also be expressed semantically if the system identifies a specific area such as a room or on a larger scale, the system can adapt to the country, the city or the time zone in which it is.

### 2.2.4 Activity

The fourth factor is the activity itself. At a micro level, it includes task characteristics and at a macro level it includes the general activity and the user's goal.

Task characteristics: For this sub-factor, the task is considered on its own. The frequency, complexity, dangerousness and confidentiality character of the task can be taken into account to adapt the system. In certain systems, icons and fast links can be added to enable easy access to frequent tasks. Extra warning messages, restoration points and password checking can be used for complex, dangerous or confidential tasks.

Task flow: Here, the task is considered as a part of a tasks flow. The tasks done before, the tasks that will be done after and the tasks done at the same time linked by dependencies can lead to different adaptations of the system. For example, if the user usually does task B after task A, the system might set a quick or automatic launch to task B each time task A is done.

User's goal: The tasks are considered as a whole forming an activity. For each activity, the user can have a different set of goals. For example, when a user is working on Photoshop®, he might be editing the photos, looking at a slide show, sorting the photos or even selecting the best photos to be printed. In all of these cases the goal is determined by the response to the question "Why is the user using the system and what does the user actually want to achieve?".

General activity: If we adopt a more global vision, the general nature of the activity weighs heavily in the successful adaptation of the system. Someone wanting to have fun, for example, will not have the same way of using a system as someone who wants to learn or work. Another dimension that deserves attention is whether the activity is engaged in, on a voluntary basis or out of obligation.

### 2.3 What aspects of the system are Adaptable?

Many aspects of applications can be adapted. We characterize these aspects mainly using the common approach of HCI engineering, PAC (presentation, abstract, control), used by Coutaz [12]. This model has the advantage of clearly separating the functional aspect of the system called "abstraction" from the interface components called "presentation". The "control" is in charge of linking these two worlds and thus externalizing the means and rules of communication.

In the next section, we clarify these aspects by using an example of a GPS system.

### 2.3.1 Abstraction

In this part we will be talking about the adaptation of the abstraction, in other words, of the information and the data proposed by the system and the way the different services behave.

Data & information: Adapting this aspect means changing the information and the data used by the system. Let us use our example to illustrate this type of adaptation: a GPS system in a car will give different information when asked for the hotels in the surrounding area. The hotels proposed will depend on the localization of the car.

<u>Service behavior</u>: The services proposed by a system need to adapt to certain circumstances. A company time-table planner for example will authorize the boss to take holidays whenever he wants but will send an approval email and mark the holidays as "to be confirmed" for any other employee.

### 2.3.2 Control

In this part we will describe the different adaptations that can be applied to the control of a system. The control module is in charge of giving access to the services and data available in the system by interacting in different ways with the user.

<u>Filtering services and data</u>: For various reasons, adaptation might mean limiting the number of services offered or providing only a partial access to a complex service. On our GPS system, for instance, the services to find a tourist attraction are only available when the system is set on "vacation" mode.

<u>Technical choice of interaction</u>: Systems can choose to accept input via many physical devices or by voice or movement recognition. The output can be delivered on screens or loudspeakers. Let us illustrate this type of aspect with our GPS system again: When the car is running, the information on the screen is read out loud by a voice synthesizer.

### 2.3.3 Presentation

In this part we will discuss the different ways the presentation of a system can be adapted.

<u>Spacio-temporal organization</u>: The elements of information can be arranged in a variety of ways. For example, the GPS system will present the descriptions of the hotels in a specific order by calculating the distance to the hotel or by taking into account the level of comfort wanted.

<u>Presentation aspects</u>: Finally, we get to the outermost layer of the surface, which includes elements such as colors, shapes, as well as the interactive elements such as buttons, boxes, menus... For example, our GPS system will change the colors and the brightness of the screen when night falls. The volume and the type of the voice used can also be adapted for different needs.

## 3. Characterizing and Quantifying Tool

As we have mentioned in the introduction, the AnAmeter tool characterizes the adaptation and measures the quantity of this adaptation. It is important to keep in mind that is does not yet measure the satisfaction of the user or the efficiency of the adaptation. In order to build a system widely accepted as a standard evaluating tool it is necessary to provide an open system that has a strong basis to support an iterative and participative building approach. We therefore present AnAmeter as a starting point for developing such a quality scale.

### 3.1 Characterizing adaptation

Using the classification presented in the previous section, we can build a first characterizing grid by crossing the adaptable aspects versus the factors of adaptation. This grid can be used to break down types of adaptability as well as the types of adaptivity (Fig. 2). Each factor (respectively aspect) is divided into sub-factors (respectively sub-aspects). The sub-factors and sub-aspects are also broken down into elements. For ease of presentation, we have not drawn these last subdivisions but each cell of the main grid contains a smaller grid composed of these elements which refer to the finest grain of description. Each lower level cell corresponds to the question "Does this aspect adapt to this factor?" If this is the case then the cell should be checked. For example, the system tested in Fig. 3 adapts the "size of the text" and the "type and color of the background" to the users "myopia".

Some of these questions might not make very much sense in certain situations or for a specific type of system. This is why we add a N/A (non-applicable) option. The N/A box can be checked if the adaptation does not seem logical or if it is not a desirable adaptation for the system. For example, an evaluating system such as the BULATS test (http://www.bulats.org/), meant to be used in a closed environment without access to internet or other software systems does not really need to be adaptable to the connectivity of the platform. The cells corresponding to adaptations to connectivity are therefore considered N/A. AnAmeter will take this into account when calculating the mean adaptation degree by ignoring these cells.

The complete evaluation requires filling out the grid and therefore answering a long list of questions. In order to ease the work of the evaluator and speed up the process, we have built an online tool for handling the grid that only requires the evaluator to check boxes. The tool is available online at http://liesp.insa-lyon.fr/AnAmeter. For the first evaluations we carried out on four well known systems, filling out the grid took about 60min.

**Fig. 2** : Characterization grid v1.0. with an example of the main grid and a smaller grid containing aspect and factor elements. To fully test a system, 2 grids like this have to be filled in, one for the system's *adaptability degree* and one for the system's *adaptivity degree*.

### 3.2 Quantifying adaptation

Now that we have built a grid to *characterize* the adaptability and the adaptivity of a system, we want to *quantify* these adaptations.

Once each cell of the smaller grid relevant to sub-aspect B and sub-factor C is filled in, an adaptation degree $A_{B/C}$ ranging from 0 to 3 is automatically calculated according to the number and distribution of the boxes checked using the rules detailed in Table 1.

For example, Fig. 2 shows the small grid of the sub-aspect "presentation aspects" and the sub-factor "perceptual/motor abilities". Once the evaluator clicks on the OK button, the adaptation degree $A_{\text{presentation aspects / perceptual, motor abilities}}$ will be automatically calculated according to the number and position of the ticks entered in the grid.

**Table 1** : Scoring process for the adaptation degree.

| Degree | Meaning | Reading in the grid | Example |
|---|---|---|---|
| $A_{B/C}= 0$ | The system does not have this type of adaptation. | No checked boxes. | 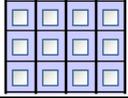 |
| $A_{B/C}= 1$ | One aspect is adapted to one factor. | One checked box. | 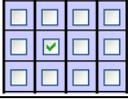 |
| $A_{B/C}= 2$ | One aspect is adapted to several factors or several aspects are adapted to one factor. | Checked boxes only on one row or only on one column. | 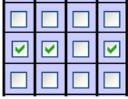 |
| $A_{B/C}= 3$ | More than two aspects adapt to more then two factors. | Checked boxes on at least two rows and two columns. | 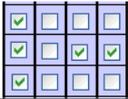 |

When all the cells in the main grid relevant to the aspect B and the factor C are filled in with a score of 0, 1, 2 or 3, a local adaptation degree, $LA_{B/C}$ is determined by calculating the average of these scores. The N/A cells will not be considered in the calculations. The results is then converted into a percentage as shown in equation n°1 (100% corresponds to a score of 3 in all the cells).

$$LA_{B/C} = \frac{\sum A_{B.item/C.item} \times 100}{(n-m) \times 3} \qquad \begin{array}{l} \text{n cells relevant to B and C} \\ \text{m N/A cells relevant to B and C} \end{array} \qquad (1)$$

One all the local adaptation degrees $A_{B/j}$ relevant to an aspect B are calculated, the semi-global *aspect* adaptation degree $AA_B$ can be found with equation n°2. In the same way, equation n°3 is used to determine the semi-global *factor* adaptation degree $FA_c$ relevant to the factor C.

$$AA_B = \frac{\sum LA_{B/j}}{n} \qquad \forall j \subset \{\text{factors}\} \qquad (2)$$

$$FA_C = \frac{\sum LA_{i/C}}{n} \qquad \forall i \subset \{\text{aspects}\} \qquad (3)$$

Finally, the global adaptation degree, GA, is determined by taking the average of the semi-global adaptation degrees - either of all the aspects or of all the factors - as shown in equation n°4.

$$GA = \frac{\sum AA_i}{n} = \frac{\sum FA_j}{n} \qquad \begin{array}{l} \forall j \subset \{\text{factors}\} \\ \forall i \subset \{\text{aspects}\} \end{array} \qquad (4)$$

To enable easy understanding of these adaptation degrees, we then identify the aspects and the factors by using the first letter of their name. Also, the adaptation degree relevant to *adaptivity* (self-adaptive) will be marked with an apostrophe (LA'$_{C/A}$, GA'…).

| | Presentation | Control | Abstraction | |
|---|---|---|---|---|
| User | 33.33 % | 37.5 % | 25 % | 31.94 % |
| Platform | 33.33 % | 33.33 % | 16.67 % | 27.78 % |
| Environment | 8.33 % | 4.17 % | 0 % | 4.17 % |
| Activity | 8.33 % | 33.33 % | 37.5 % | 26.39 % |
| | 20.83 % | 27.08 % | 19.79 % | 19.79 % |

- **Semi-global** adaptation to the Platform: 27.78 %
- **Semi-global** Presentation adaptation: 20.83 %
- **Local** adaptation of the Control to the Activity: 33.33 %
- **Global** adaptation of the system: 19.79 %

**Fig. 3** : Example of local, semi-global and global adaptation degrees.

Fig. 3 illustrates these equations:
Local adaptation of the control to the activity: $LA_{C/A}$ = 33.33 %
Adaptation of the presentation aspect: AAp = 20.83 %
Adaptation to the platform factor: FAp = 27.78 %
Global adaptation of the system: GA = 19.79 %

AnAmeter provides an overall score for the adaptation degree of the system yet also provides sub-scores (local and semi-global). This is very useful for systems that are specialized in a certain type of adaptation.

## 4. Discussion

Our proposal for characterizing adaptation is to use a scoring matrix in order to quantify local, semi-global and global adaptation degrees of a system. The advantages of this tool are:

1) Its simplicity. The user fills out the grid by answering simple Boolean questions of the following type: does a precise aspect of the system adapt to a precise factor? Clear examples with references are available for each type of adaptation. The resulting grid and charts are automatically generated.

2) Its precision. The tool provides precise local evaluations. This is very useful for specialized systems that concentrate on one aspect of adaptation. AnAmeter also allows for the possibility of evaluating the system from two fundamental different points of view: adaptability (user-initiated adaptation) or adaptivity (automatic system-initiated adaptation).

3) Its ease for comparing adaptive systems. The final result grid can be compared to

any other result grid. In addition, AnAmeter has already been used to evaluate the adaptation degrees of four well-known systems. These evaluations available on the web platform can therefore be used as a basis for further comparisons.

4) Its extensibility and flexibility. Our idea was to offer a robust basis for the community to build on. The architecture of our tool makes it easy to extend by adding other elements or refine it by dividing the sub-categories or extending the measuring scale to 0-5 or 0-10.

5) Its accessibility. AnAmeter is freely accessible on the web along with a selection of completely tested systems and more than 300 examples of adaptation types (http://liesp.insa-lyon.fr/AnAmeter). This makes it possible for the same system to be tested by several evaluators who could then combine their results to obtain a mean value for the adaptation degree.

Although AnAmeter has many advantages, the fact that the approach tries to be as complete as possible extends the time required to evaluate a system to approximately one hour. Indeed this first version of the grid contains 22 aspect elements and 59 factor elements which adds up to more than a 1000 Boolean questions to answer for a highly adaptive or adaptable system. Of course, for most of the systems, entire sections of the grid will be left out or marked as non-applicable, greatly reducing the amount a work. By creating an online tool that enables easy manipulation of the grid and calculates the adaptation degree automatically, we have lightened the task but it is still represents quite an investment of time and effort. We hope it will be possible to improve the grid with the scores and the comments of people who use it.

## 5. Future work

We believe that building an evaluating tool, widely accepted by the community can only be done in a cooperative way with the help of the members of this community. AnAmeter was created to serve as a basis for building on and this is why we created an open, extensible and flexible online framework.

In the near future, we plane on adding a subjective measure of the user's satisfaction of the adaptation to establish a global evaluation mark as seen in the first section (see Fig. 1) In order to do this, AnAmeter can be reconfigured so that, instead of indicating if the adaptation is available or not by checking the boxes, the evaluator will enter a measurement of his or her satisfaction.

Another Idea is to ask people with different user profiles to test the same system in order to see if the results coincide or not. This will allow us to measure the reliability of AnAmeter.

If the system is fully developed, the designer might also want to ask a user to evaluate the system with AnAmeter to see if he is aware of all the adaptable (user-initiated) options and if the adaptive dimensions (automatic system-initiated) have been noticed.

The next step is to test the AnAmeter tool for ease of use by asking other people to use it to evaluate systems on their own and send us feedback. Now that AnAmeter is available on the web is should be easy to launch an evaluation campaign. A simple questionnaire could be added to know if the evaluators found the system clear and easy to use.

## 6. Conclusion

In this paper, we present AnAmeter, a tool to characterize the multiple facets of adaptability and a quantification technique to measure the adaptability degree of an interactive system. We discuss the multiple facets of adaptation, primarily the aspects and factors of adaptation that serve as parameters. Then, we suggest the use of a scoring matrix to evaluate local, semi-global and global adaptation of an interactive system. We provide a first version of the scoring technique and simple formulas for calculating these adaptation degrees. The AnAmeter tool is presented as a starting point for the community to cooperatively build a widely accepted framework for evaluating any kind of adaptable or adaptive systems.